# Limitations of long term stability in a coherent population trapping Cs clock

Olga Kozlova[1], Jean-Marie Danet, Stéphane Guérandel, and Emeric de Clercq

LNE-SYRTE, Observatoire de Paris, CNRS, UPMC, 61 Avenue de l'Observatoire, 75014 Paris, France.
[1] present adress : Laboratoire Commun de Métrologie, LNE-CNAM, 61 rue du Landy, 93210 La Plaine Saint Denis, France.

*Abstract*— Vapor cell atomic clocks exhibit reduced frequency stability for averaging time between about one hundred and a few thousand seconds. Here we report a study on the impact of the main parameters on the mid-to-long term instability of a buffer-gas vapor cell Cs clock, based on coherent population trapping (CPT). The CPT signal is observed on the Cs $D_1$ line transmission, using a double $\Lambda$ scheme and a Ramsey interrogation technique. The effects on the clock frequency of the magnetic field, the cell temperature, and the laser intensities are reported. We show in particular that the laser intensity shift is temperature dependent. Along with the laser intensity ratio and laser polarization properties, this is one of the most important parameters.

## I. Introduction

Traditional rubidium gas cell frequency standards [1, 2] are attractive for their small size and moderate power consumption combined with a good short term stability, ranging in fractional value from $10^{-11}\tau^{-1/2}$ to $2\times10^{-12}\tau^{-1/2}$ for the best spatial prototypes, where $\tau$ is the averaging time. They are thus widely-used in applications such as laboratory instrumentation, telecommunication, onboard satellites and global navigation satellites system (GNSS), etc. Their ubiquity is however limited by their main drawback, a frequency drift of the order of $\pm\ 10^{-13}$/day in fractional units [3], which is poorly explained until now. Several effects, depending on time or on environmental parameters such as temperature, can be put forward to explain this drift, mainly related to the buffer gas pressure shift, the light shift, the microwave power, or a mixture of these [3-6].

A great deal of work has been devoted to the search for alternatives to the usual pumping scheme using a discharge lamp and a filter cell, and for improvement of the short-term frequency stability. The replacement of the lamp by a diode laser gave great hope [7, 8, 9]. Very recently noteworthy results were obtained by Bandi et al. [10] with short term stability better than $3\times10^{-13}\tau^{-1/2}$. Different ways have been explored by Godone's group in INRIM. A vapor cell pulsed optically pumped rubidium maser has shown a stability of $1.2\times10^{-12}\tau^{-1/2}$ [11]. Using pulsed laser pumping and a Ramsey microwave interrogation on a Rb cell, the same group reported an outstanding fractional stability of $1.7\times10^{-13}\tau^{-1/2}$ [12]. Note that the use of a pumping laser also allows Cs vapor cell clocks.

The microwave interrogation can be replaced by the coherent population trapping technique (CPT) [13]. In this case there is no microwave cavity; the microwave being carried by the frequency difference between two laser beams [14, 15]. CPT allows the development of miniaturized atomic clocks, first developed at NIST [16], with short-term stability of $4\times10^{-11}\tau^{-1/2}$ [17]. For large scale systems a stability of $1.4\times10^{-12}\tau^{-1/2}$ was demonstrated by Zhu [18]. Godone et al. reported a stability of $3\times10^{-12}\tau^{-1/2}$ with a CPT vapor cell Rb maser [19]. At SYRTE we have developed a CPT Cs cell clock using a Ramsey interrogation [20] which exhibits a $7\times10^{-13}\tau^{-1/2}$ stability [21].

All these new results show a noticeable improvement of the short term stability. For short averaging time the stability, measured by the Allan deviation decreases as $\tau^{-1/2}$, characteristic of a white frequency noise. Nevertheless for all these vapor cell clocks, whatever their scheme, the medium-to-long-term stability increases for averaging times typically larger than 100 s to a few $10^3$ s. The long-term stability problem is still open. Effects related to the pumping lamp and isotopic filter are eliminated; however, those related to temperature, pressure shift and light shift [22, 23], although different, remain, while new sources of instability are possible. The comparison between standards of several different realizations could lead to a better understanding of the phenomena involved.

The medium-term stability of our prototype is limited to the $2\text{-}3\times10^{-14}$ level for an averaging time of about 4 000s. We have investigated some of the main sources of frequency shift which could affect the medium-to-long term stability of our clock prototype and could explain its observed limit. The experimental set-up is briefly described in the next section. An example of the measured frequency stability is shown. The sensitivity to the magnetic field is investigated in Section III. We show that the double $\Lambda$ scheme used induces new features of the magnetic shift compared to the well known quadratic Zeeman shift. Section IV is devoted to the effect of the collisions between Cs atoms and buffer gas molecules. A mixture of buffer gases, optimizing the signal and the temperature sensitivity, is determined. The experimental study of the impact of the laser intensities on the clock frequency is reported in Section V. The shift rate is shown to be dependent on the cell temperature and larger for a variation of the intensity ratio than of the total intensity.

## II. EXPERIMENTAL SET-UP

The experimental set-up is shown in Fig. 1. In order to pump the atoms in the clock dark state (linear superposition of $|F = 3, m_F = 0\rangle$ and $|F = 4, m_F = 0\rangle$) we use two external cavity diode lasers tuned to the Cs $D_1$ line at 894 nm. The master laser is frequency locked to the ($F=4 - F'=4$) transition via a saturated absorption scheme in an auxiliary evacuated Cs cell. The scheme contains a frequency modulated acousto-optic modulator (AOM), not shown in Fig.1, in order to avoid modulating the master laser frequency. It allows also shifting the laser frequency compared to the Cs reference for compensating the buffer gas shift in the clock cell. The slave laser is phase locked to the master laser with a frequency offset tunable near 9.192 GHz, i.e. around the ($F=3 - F'=4$) transition, by comparison with a low-noise synthesized microwave signal [24] driven by a hydrogen maser. For this purpose, the two laser beams are superimposed by means of the polarizing beam splitter cube (PBS) and are detected by the fast photodiode (PD0).

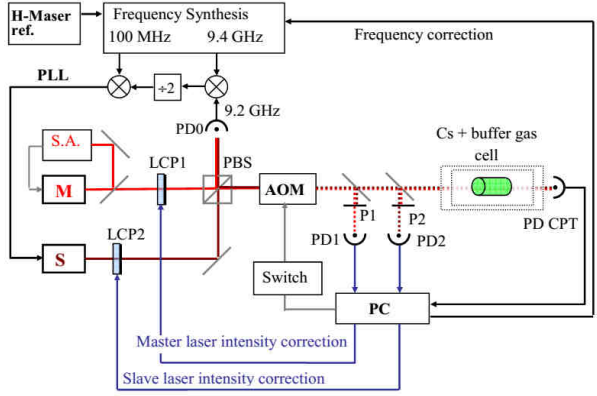

Fig. 1. CPT pulsed clock scheme. The explanations are given in the text. M – master laser; S – slave laser; S.A. – saturated absorption scheme; LCP1 and LCP2 – liquid crystal variable retarder with polarizer; PBS – polarizing beam splitter cube; PD0 – fast photodiode; PLL – phase-lock loop; AOM – acousto-optic modulator; P1 and P2 – polarizers; PD1, PD2 and PD CPT – photodiodes; PC – computer.

We use a so-called double Λ scheme (see Fig. 2) with two linear and orthogonally polarized laser beams (lin-perp-lin configuration [20]) propagating parallel to the static magnetic field applied to the cell. In order to operate in the pulsed mode (Ramsey interrogation), a pulse sequence is applied to the acousto-optic modulator. The typical sequence consist of a train of 2-4 ms duration pulses separated by a free evolution time $T_R = 4$ ms, i.e. about the hyperfine coherence relaxation time [25]. In order to avoid temperature dependent birefringence effects in the AOM crystal, its temperature is regulated to within several mK. The laser intensity transmitted through the Cs cell is measured by the photodiode (PD CPT), linked to the computer (PC), which locks the synthesizer frequency to the clock resonance. The laser intensities of the master and slave lasers are detected separately by the photodiodes PD1 and PD2. They can be controlled using voltage-controlled liquid crystal variable retarders LCP1 and LCP2 (Meadowlark Optics). The lock-in of the intensity is integrated in the software, which also controls the CPT frequency lock and the pulse sequence.

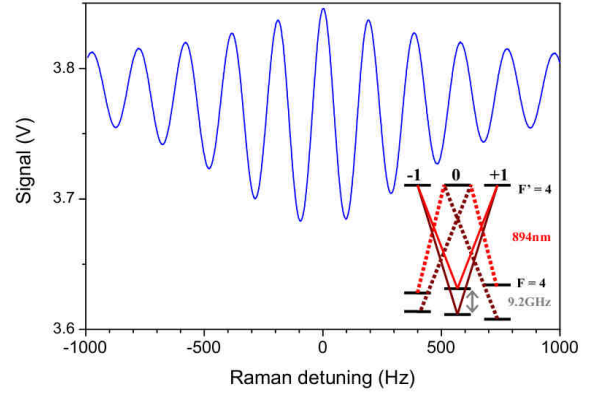

Fig. 2. Measured Ramsey fringes. Inset: double Λ scheme. The scheme is similar to a $\sigma^+$, $\sigma^-$ excitation. Only the clock levels ($m_F = 0$), and the nearest Zeeman sublevels are shown. $F = 3$ and $F = 4$ are the hyperfine levels of the ground state, $F'$ is one excited level of the $D_1$ line. The clock transition is induced by the transitions shown in solid lines. The transitions in dashed lines induce hyperfine transitions with $\Delta m_F = 2$.

The Cs cell is made of Borofloat 33, a borosilicate glass made by Schott Company, with a diameter of 2 cm and length 5 cm, filled in the laboratory with an optimized Ar-$N_2$ buffer gas mixture (see Section IV). Its temperature is regulated at the mK level. The cell is placed inside a solenoid, surrounded by two mu-metal magnetic shields. Special attention was paid to reduce the influence of spurious magnetic fields. The cell mount is made in nonmagnetic copper. The thermistors, used to control the cell temperature, are also nonmagnetic. A coaxial nonmagnetic heating wire is used; the winding is made in such manner that each step is wound in opposition to the previous in order to reduce the spurious field created. Thanks to these precautions we estimate that the contribution of the fluctuations of residual magnetic field in the cell to the clock frequency instability is negligible compared to other sources. The current source of the solenoid is a home-made modified diode laser current source.

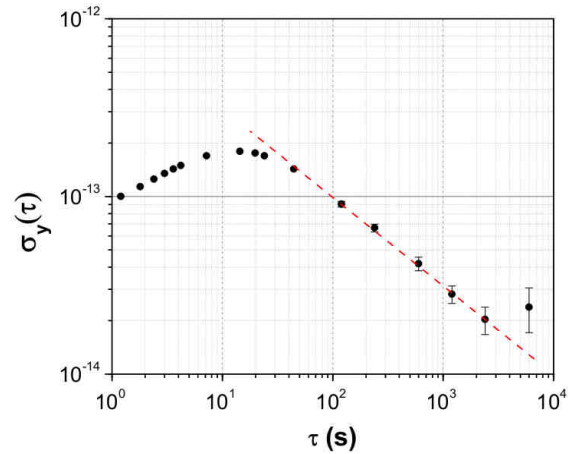

Fig. 3. Allan standard deviation of the clock as a function of the averaging time. Dots: experimental values measured versus a hydrogen maser, dashed line: fit with a $1 \times 10^{-12} \tau^{-1/2}$ line.

An example of the fractional frequency stability measured versus the hydrogen maser is shown on Fig. 3. For averaging times $\tau < 20$ s the Allan standard deviation is determined by the servo-loop which was not optimized for short times. For $20$ s $< \tau < 2000$ s the Allan standard deviation decreases as $1 \times 10^{-12} \tau^{-1/2}$ characteristics of a white frequency noise; it increases for larger times.

### III. MAGNETIC FIELD AND POLARIZATION EFFECTS

The main source of drift of the clock frequency induced by the magnetic field is the quadratic Zeeman shift of the clock transition [2]:

$$\nu_z = 0.0427453 B^2, \qquad (1)$$

where $\nu_Z$ is expressed in Hz, and $B$ in µT. However the fringe amplitude is also dependent on the magnetic field. It oscillates as a function of the magnetic field value with a period related to the fringe width [21] as shown in Fig. 4a.

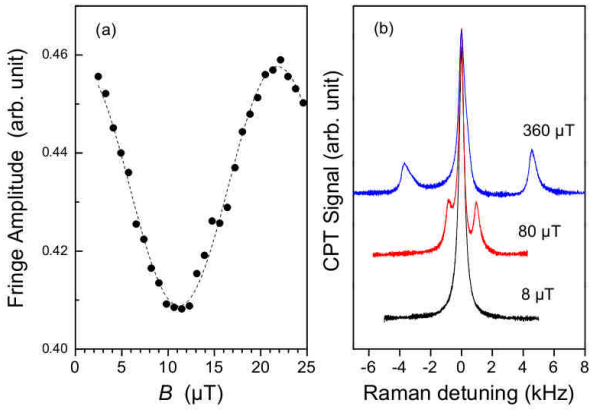

Fig. 4. (a) Amplitude of the central Ramsey fringe as a function of the static magnetic field $B$. The dots are the experimental data, the dashed line is a fit with a cosine function. $T_R = 4$ ms. (b) CPT signal with a cw interrogation for three values of the magnetic field. The resonance is split in three resonances at high magnetic field. For better visibility purpose a vertical offset is applied to each curve, and the curves are centered on the 0-0 resonance frequency at low field.

This oscillation is explained by the superimposition of three fringe systems, the clock signal, ($m_F = 0$ - $m_F = 0$) transition, and the fringes corresponding to ($m_F = -1$ - $m_F = +1$) or $\Delta m_F = 2$ transitions, whose resonance frequencies are very close to the clock transition and which move apart when the magnetic field increases. The three peaks can be seen at high magnetic field with a continuous (CW) interrogation (Fig. 4b). The neighboring resonances split the clock resonance with frequencies given by the Breit-Rabi formula [26]:

$$\nu_{z,\pm} = \pm 11.1649 B - 0.002672 B^2, \qquad (2)$$

where $\nu_{z,\pm}$ (Hz) refers to the transitions ($F = 3$, $m_F = +1$) - ($F = 4$, $m_F = -1$) and ($F = 3$, $m_F = -1$) - ($F = 4$, $m_F = +1$), respectively; $B$ is in µT. It is worth noting that the frequencies $\nu_{z,\pm}$ are the frequency differences between these transitions and the clock transition.

The fractional stability of the magnetic field strength shown in Fig. 5 was recorded with the $B$ value yielding the maximum fringe amplitude, $B \approx 22$ µT. The stability of $B$ is measured by means of the frequency stability of the clock locked to the transition ($\Delta F = 1$, $m_F = 1$) which deviates of the clock transition ($\Delta F = 1$, $m_F = 0$ - $m_F = 0$) from 7.008 kHz/µT, neglecting the small quadratic term.

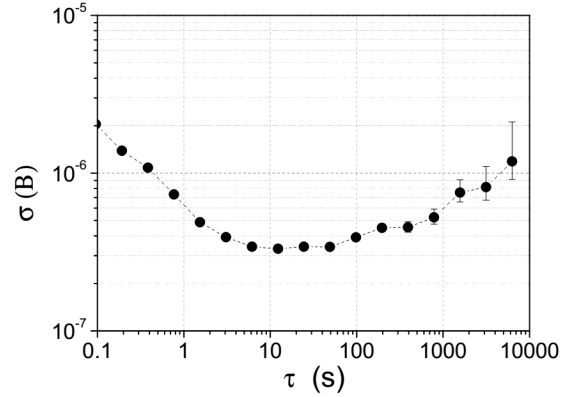

Fig. 5. Fractional Allan standard deviation of the static magnetic field strength. The dashed line is only a guide for the eye.

According to (1) a fractional fluctuation of $B$ of $2 \times 10^{-6}$ leads to a fractional frequency shift of $0.9 \times 10^{-14}$. However this estimation does not take into account the influence of the superimposed ($\Delta m_F = 2$) transitions, which is analogous to the well-known Rabi pulling effect [2]. A fluctuation of $B$ entails unequal shifts of these transitions in opposite directions. We have performed a rough estimation of the resulting shift of the maximum of the signal. In a first approximation the observed signal can be considered as the sum of three cosine functions of period two linewidths $w$, and of amplitudes 1, $a_+$, $a_-$, for the 0-0, ($F = 3$, $m_F = +1$)- ($F = 4$, $m_F = -1$) and ($F = 3$, $m_F = -1$)- ($F = 4$, $m_F = +1$) transitions, respectively. A power series expansion of the first order around $\nu_Z$ allows us to get an analytical expression of the observed Zeeman shift:

$$\nu_z' \approx \nu_z + \frac{w}{\pi} \frac{a_+ \sin(\pi \nu_+ / w) + a_- \sin(\pi \nu_- / w)}{1 + a_+ \cos(\pi \nu_+ / w) + a_- \cos(\pi \nu_- / w)}. \qquad (3)$$

For laser beams polarized perfectly linearly and orthogonally the amplitudes $a_+$, $a_-$ are equals. However birefringent effects are unavoidable, arising from the cell windows, the mirrors or the AOM. A weak ellipticity of one beam polarization is enough to impact the equilibrium of $a_+$, $a_-$ increasing the difference $\nu_z' - \nu_z$. Fig.6a shows the CW CPT signal for different elliptical polarization of the laser beams achieved by setting a retardation plate. The signal is recorded at high magnetic field (60 µT) so that the peaks are well separated. The measured and calculated resulting shifts are shown in Fig. 6b for a Ramsey interrogation and in the case of the largest unbalance of Fig. 6a ($a_+ = 0.17$, $a_- = 0.03$). The calculated shift shows a magnification of the oscillations compared to the experimental shift. The difference between both curves could be explained by a difference of the values of $a_+$ and $a_-$ measured with a CW interrogation at high magnetic field and their values with a Ramsey interrogation at lower field used when the frequency is locked. Indeed, under Ramsey interrogation (pulsed mode) the AOM birefringence differs from the one in CW mode because of thermal effects in the AOM crystal. It is also possible that the amplitudes measured in CW mode correspond to the steady-state, while in the pulsed mode the steady-state is not reached. The approximations used in (3) are also possibly too rough. Nevertheless, using the amplitudes $a_+$, $a_-$ as free parameters, the experimental shift is well fitted by (3).

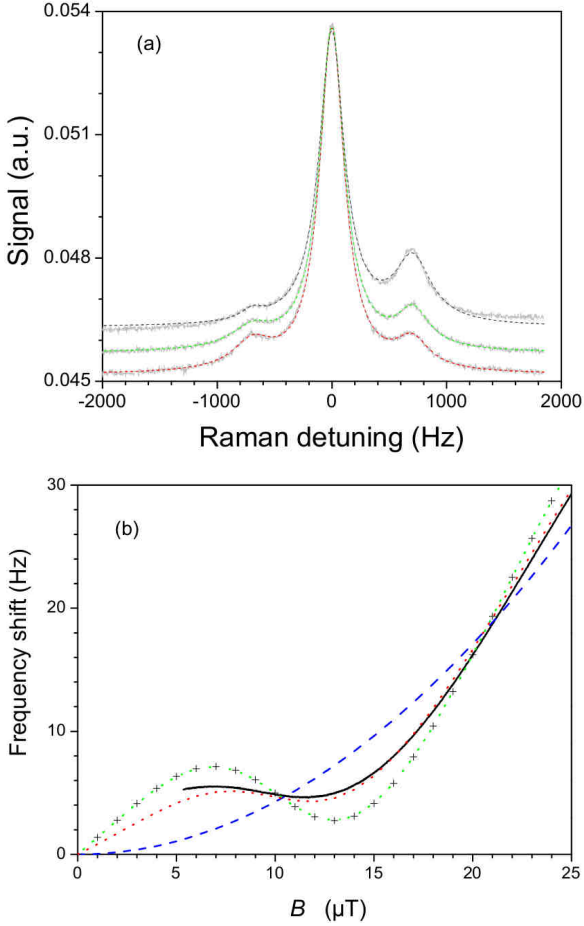

Fig. 6. (a). CPT signal with a CW interrogation for three different polarisation states. $B = 60$ µT. Lower curve : linearly and orthogonally polarized beams. Middle curve: the beams are slightly elliptically polarized, upper curve: same as middle curve with an ellipticity increased. Each curve is fitted with the sum of three Lorentzian profiles of adjustable fractional amplitudes. (b). Calculated shift of the observed clock frequency as a function of the applied magnetic field. Dashed line: theoretical shift of the 0-0 transition, solid line: experimental data, dashed line with crosses: shift calculated with (3) ($a_+$ = 0.17, $a_-$ =0.03) , dotted line: fit of (3) with $a_+$ and $a_-$ as free parameters, ($a_+$ = 0.14, $a_-$ =0.06).

At our working point, the maximum of the fringe amplitude, the shift is unchanged compared to the one of (1). However the slope of the shift versus the magnetic field value is different. The amplitudes $a_+$, $a_-$ are estimated to 0.22 and 0.14. A $2\times10^{-6}$ fractional fluctuation of $B$ will now lead to a fractional frequency shift of $1\times10^{-14}$. It is worth noting that the sensitivity to the magnetic field (slope of the solid line curve in Fig. 6b) cancels in two points (around 7 and 12 µT in Fig. 6b). It would be interesting to fix the static magnetic field strength at one of these values when the stability of the current source is reduced, or for long term stability purposes. Note that with equal amplitudes, $a_+ = a_-$, the effect is still present but much weaker; in this case the sensitivity does not cancel.

## IV. Cs - BUFFER GAS PRESSURE SHIFT

There is a common technique for alkali vapor cell atomic clocks gas to add a buffer gas in order to reduce the relaxation by collisions on the cell walls and to reduce the Doppler broadening by the Dicke effect [2]. However the collisions between alkali atoms and the buffer gas induce a temperature dependent frequency shift of the clock transition [2]. The temperature dependence can be cancelled around a chosen temperature using a mixture of buffer gases with opposite signs of temperature coefficients [27], like $N_2$ and Ar with Cs. In this case the pressure shift can be written [2]:

$$\upsilon_T = \frac{P_0}{1+r} \times \left( \beta_{N_2} + r\beta_{Ar} + \right. \\ \left. (\delta_{N_2} + r\delta_{Ar})(T-T_0) + (\gamma_{N_2} + r\gamma_{Ar})(T-T_0)^2 \right), \quad (4)$$

where $P_0$ is the buffer gas pressure in the cell at the reference temperature $T_0$, $T$ is the cell temperature, $r$ is the pressure ratio $P_{Ar}/P_{N2}$. The coefficients $\beta_{N2}$, $\beta_{Ar}$ (Hz torr$^{-1}$), $\delta_{N2}$, $\delta_{Ar}$ (Hz torr$^{-1}$ K$^{-1}$), $\gamma_{N2}$, $\gamma_{Ar}$ (Hz torr$^{-1}$ K$^{-2}$) are the coefficients measured at $T_0$ for $N_2$ and Ar, respectively.

Thanks to our previous investigation on the Cs-buffer gas collisional shift [28] we can calculate the optimal buffer gas mixture cancelling the temperature dependence around $T_i$, the chosen inversion temperature:

$$r = -\left(\delta_{N_2} + 2\gamma_{N_2}(T_i - T_0)\right)/\delta_{A_r} \\ \approx 0.724(12) - 0.00441(9)(T_i - T_0), \quad (5)$$

the numbers in parentheses are the uncertainties, $\gamma_{Ar}$ is null or very weak and can be neglected, $T_0$ =0°C, $T_i$ in °C. The working temperature is chosen in order to get the maximum amplitude of the CPT signal which is expected to be around 29-30°C for this size of cell and pressure range, leading to $r$ = 0.60.

The filling pressure of the buffer gas was calculated for maximizing the hyperfine coherence relaxation time. The relaxation rate of the coherence is mainly determined by the sum of three terms [2]: collision rate on the cell walls, decreasing when $P_0$ increases, the collisions with buffer gas, increasing when $P_0$ increases, and the spin exchange rate between Cs atoms, which depends only on the density of the Cs atoms and not of $P_0$. With the values of the diffusion coefficients and cross-section listed in [2] the optimum pressure value for the ($r$ = 0.6) N2-Ar mixture is $P_0$ = 20.5 torr.

A Cs cell was filled in the laboratory with theses specifications. The measured relaxation time (4.7 ms at 29°C) is about half the computed value (11 ms). This discrepancy is unexplained at present time. Fig 7 shows the clock signal as a function of the cell temperature. The signal reaches a maximum at 29°C, after which the medium becomes optically thick [29]. However the temperature of the maximum amplitude also slightly depends on the laser intensity, see inset in Fig. 7.

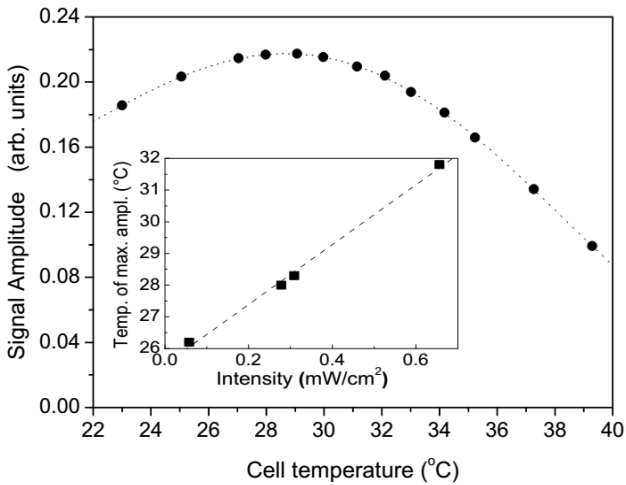

Fig. 7. CPT signal amplitude as a function of cell temperature. Dots: experimental data; the dashed line is only a guide for the eye. Ramsey interrogation: pulse duration 2 ms, free time evolution 4 ms. Total laser intensity: 0.3 mW/cm$^2$. Beam diameter: 11 mm. Inset: dependence of the temperature of the maximum amplitude on laser intensity.

The frequency shift measured as a function of the cell temperature is shown in Fig. 8. The frequency shift is corrected for the magnetic shift and extrapolated to null laser intensity.

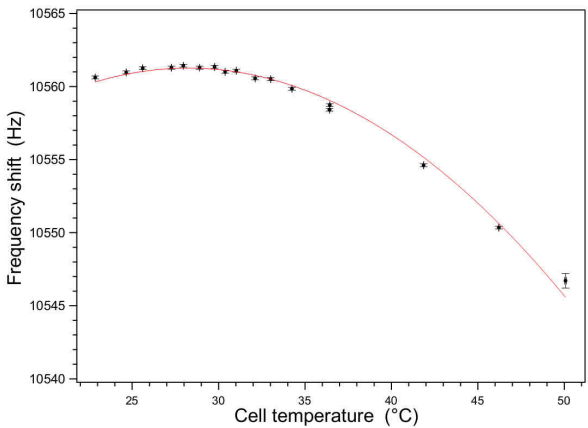

Fig. 8 CPT clock resonance frequency shift in the buffer-gas cell as a function of the cell temperature. Buffer gas mixture: Ar-N$_2$, total pressure 21 torr, ratio r(Ar/N$_2$)=0.6. Dots: experimental values with their error bars, the solid line is a fit by (4).

The temperature shift is fitted with (4), the free parameters are the total pressure $P_0$ and the pressure ratio $r$. The fitted values are $P_0$ = 20.95(4) torr, $r$ = 0.600(3). The inversion temperature is 28.2°C. As the temperature of the cell is regulated better than 1 mK at 29°C for an averaging time of $10^4$ s, the contribution of the pressure shift to the frequency stability is then $0.6 \times 10^{-14}$ at $10^4$ s. We estimate that the effect of the temperature gradients [30] in the cell can be neglected at the present level of stability. This effect will need further investigations for future clocks with improved stability.

Note that with a shift of about 5 kHz/torr the mean frequency is very sensitive to any pressure variation. We have no evidence of drift of the Ar or N2 pressure with time, but in this case, for example by a chemical process, this could affect strongly the long term stability. The fractional frequency shift due to the He permeation in the cell has been estimated following the calculation described in [6]. It leads to an Allan standard deviation of $4 \times 10^{-15}$ for an averaging time of $10^4$ s. It is one order of magnitude lower than our present frequency stability level and we do not take it into account. But it should be considered for the clocks with improved frequency stability and for longer averaging times.

## V. LIGHT DEPENDENT EFFECTS

The effect of the laser light is known as one of the major effects which can limit the long-term stability of CPT-based clocks [31, 32]. Here we investigate three main effects: the frequency dependence on the total laser power, its sensitivity to the cell temperature, and finally the effect of a laser intensity ratio variation. The frequency shift of the clock transition is shown in Fig. 9 as a function of the total laser intensity for different cell temperatures. The shifts are measured in the Ramsey mode, the laser intensity value is the intensity measured during a pulse.

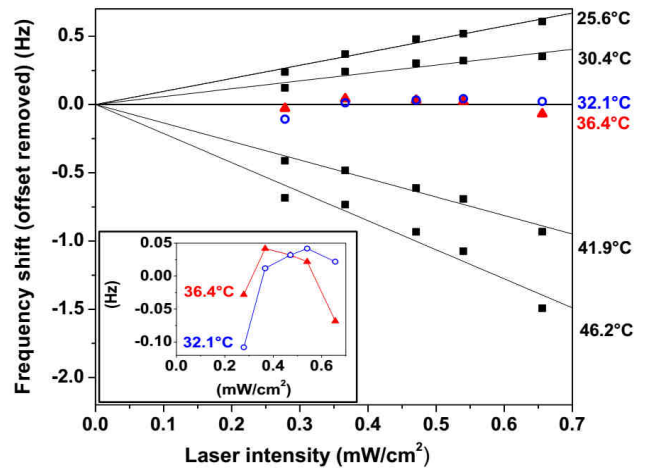

Fig.9. Frequency shift of the clock transition versus total laser intensity for different cell temperatures. Ramsey interrogation: pulse duration 2 ms, free time evolution 4 ms. For better visibility a different offset is applied for each temperature in order to compensate the pressure shift. In the inset: zoom for the temperatures: 32.1°C and 36.4°C.

For better visibility in Fig. 9 the data are corrected for the buffer gas pressure shift (Fig. 8) at each temperature value. The two laser intensities are equal within a few percent at the cell input. The optical frequency detuning is estimated to be 18 MHz, small compared to the 350 MHz homogeneous linewidth of the optical transition broadened by buffer gas collisions. The shift is clearly dependent on the cell temperature. Its slope changes with the cell temperature, and even can change sign. Such a change of sign was already reported in [32] using a continuous interrogation and a phase-modulated laser tuned to the (F'=3) level. However no sign change was observed when the laser was tuned to the (F'=4) level. Here the sign change occurs with lasers tuned to the (F'=4) level, the difference may come from a lower working temperature. The inset in Fig. 9 shows a zoom for the temperature range where the slope of the shift seems to cancel. The cancellation of the intensity sensitivity occurs only for a precise value of the couple laser intensity-cell temperature. The mean shift rates of Fig. 9, except those which change of sign with the intensity, are reported in Fig. 10 as a function of the temperature.

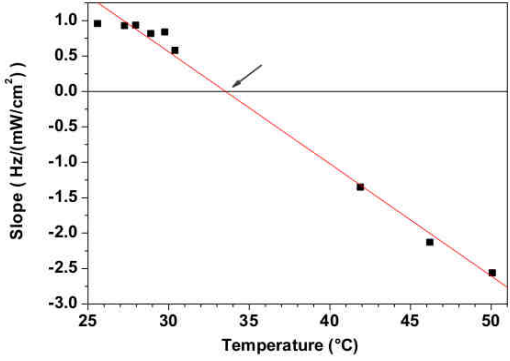

Fig. 10. Mean slope of the frequency shift versus laser intensity as a function of the cell temperature, data of Fig.9. The solid line is a linear fit. The arrow shows the temperature where the mean slope should be zero. Note that there is no possibility to define a mean slope for the measurements made in the 30°C - 40°C temperature range because of the quadratic dependence (see Fig.9).

A careful inspection of data of Fig. 9 suggests that the frequency shifts do not vary linearly with the laser intensity, as expected for a pure light shift effect [33-38]. This is confirmed by measurements at higher laser intensities, see Fig. 11.

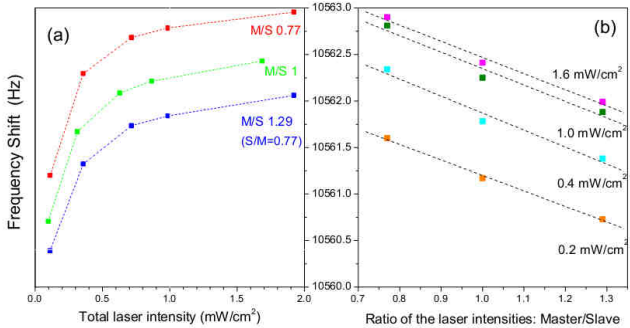

Fig. 11. a) Frequency shift as a function of the total laser intensity for three values of the intensity ratio M/S = (master laser intensity) / (slave laser intensity): 0.77, 1, and 1.29. b) Frequency shift as a function of M/S for different constant total intensities.

We have investigated the effect of an unbalance of laser powers, thanks to an independent control of both laser powers allowed by our set-up. In Fig. 11a, the measured clock frequency is shown as a function of the total laser intensity for three values of the intensity ratio *M/S*. *M/S* is defined as the master laser intensity, tuned to the (F=4 – F'= 4) transition, divided by the slave laser intensity, tuned to the (F=3 – F'= 4) transition. Fig. 11b shows the same data plotted as a function of M/S at constant total intensity. The mean shift slope -1.8Hz/(M/S) seems weakly dependent on the total intensity value.

In summary, Figs. 9, 10, 11 show that the intensity shift is temperature-dependent, it depends on the total laser intensity, but in a non-linear way, and it depends on the ratio of the intensities. A quantitative explanation of the clock frequency shift with the laser intensity or with the laser frequency is not easy because many different effects are involved. The usual light shift (AC Stark-shift) [33], computed for optically thin medium as for a two-photon transition, i.e. the shift difference of both hyperfine levels [34-36], or taking into account the specificity of CPT [37, 38], may not be the major effect. The line shape is actually distorted by other effects, temperature and intensity-dependent, which result in a frequency shift of the locked frequency [39]. It is the Rabi pulling effect [2] induced by the bottom of the linear optical absorption profile, or by the wing of the neighboring optical line (the hyperfine structure of the excited state is 1168MHz) broadened by the buffer-gas and by Doppler effect (width 360 MHz). The transitions (-1 - +1) shown in Fig. 2, are of unequal amplitudes, intensity-dependent, and are very close of the clock transition. They can induce a significant Rabi pulling. We have also to take into account inhomogeneous broadening and light intensity gradient in the cell [40], and the differential absorption of both beams along the cell [33, 39], which could explain the temperature effect. All these effects are mixed, and they have to be integrated over the broad velocity distribution. The evaluation of each one requires specific study which is out of the scope of this paper.

From a practical point of view we can introduce an "effective" light shift written in a first approximation as a linear development of first order terms.

$$\delta\nu = \left(\frac{\partial\nu}{\partial I}\right)_{T,M/S} \delta I + \left(\frac{\partial(\partial\nu/\partial I)}{\partial T}\right)_{I,M/S} I\delta T + \left(\frac{\partial\nu}{\partial(M/S)}\right)_{I,T} \delta(M/S) \qquad (6)$$

where $\partial\nu$ is the clock frequency variation, $\delta I$, $\delta T$ and $\delta(M/S)$ are the variation of the total laser intensity, the temperature and the intensity ratio M/S, respectively. $(\partial\nu/\partial I)_{T,M/S}$ is the light shift rate at the working temperature and the working ratio M/S. $(\partial(\partial\nu/\partial I)/\partial T)_{I,M/S}$ is the sensitivity to the temperature of the light shift rate, $(\partial\nu/\partial(M/S))_{I,T}$ is the sensitivity of the frequency to the laser intensity ratio M/S.

At the cell temperature of 29°C, and a laser intensity of 1mW/cm$^2$, the measured shift rate for a total laser intensity variation is 0.33 Hz/(mW/cm$^2$); the temperature sensitivity is -0.13 (Hz/(mW/cm$^2$)/°C. A 1 mK temperature variation at constant laser intensities causes a shift of $1.4\times10^{-14}$. It is worth noting that this value is of the same order of magnitude or greater than the effect of the pressure shift. For a $\delta I/I = 1\times10^{-3}$, $\delta T = 1$ mK, $\delta(M/S) = 4\times10^{-4}$, the three terms of (6) are equal to $3.6\times10^{-14}$, $1.4\times10^{-14}$, and $8\times10^{-14}$, respectively. The clock frequency is then much more sensitive to a change of one laser intensity with respect to the other one than to the total laser intensity. The shift rates are $(\partial\nu/\partial M)_{T,S}$ = -3.0 Hz/(mW/cm$^2$) and $(\partial\nu/\partial S)_{T,M}$ = +3.7 Hz/(mW/cm$^2$) for a variation of the master and slave laser intensity, respectively. Equation (6) applies in the case where the total laser intensity is regulated. In order to reduce the fluctuations of the laser intensities we locked the intensity of each component independently with the system of liquid crystal variable attenuators (see Fig 1) at a $2\times10^{-4}$ fractional level for a $4\times10^{3}$ s averaging time. The effective light shift can then be written:

$$\delta\nu = \left(\frac{\partial \nu}{\partial M}\right)_{T,S} \delta M + \left(\frac{\partial \nu}{\partial S}\right)_{T,M} \delta S + \left(\frac{\partial(\partial \nu/\partial I)}{\partial T}\right)_{I,M/S} I\delta T. \quad (7)$$

Thus the contribution to the clock frequency stability of the laser intensity noise, for a total laser intensity of 1mW/cm$^2$, is $5.2\times10^{-14}$ for uncorrelated noises (quadratic sum of both first terms of (7)), and $0.8\times10^{-14}$ for perfectly correlated fluctuations (algebraic sum of both term). The measured stability lower than $5.2\times10^{-14}$ shows that the laser intensity fluctuations are partially correlated, which can be explained by their common optical path through air turbulences and optical components.

The clock frequency depends also on the optical detuning. By varying the detuning using both AOM of the set-up we have measured a shift of 7 mHz/MHz i.e. $7.6\times10^{-13}$/MHz in fractional value. We estimate that the master laser frequency is stabilized to better than 10 kHz for $10^4$ s [40] so that the shift is smaller than $8\times10^{-15}$. The optical transition is shifted by the buffer gas of nearly 170 MHz, with a sensitivity of about -0.17 MHz/K [39]. A 1mK temperature variation leads then to a negligible shift.

## VI. DISCUSSION

In summary we have reviewed some of the main effects affecting the resonance frequency in our CPT Cs clock prototype. The resonances ($m = -1 - m = +1$) between neighboring Zeeman sublevels of the clock levels, induced by the double Λ excitation scheme, lead to a shift of the locked frequency with the magnetic field different than the second order Zeeman effect. For sufficiently large unbalance of their amplitudes, it is possible to find a field value for which the magnetic sensitivity cancels. However, this value depends on the unbalance of the (-1 – +1) resonance amplitudes which can vary with time, according to polarizations, leading to a new source of frequency instability. A 0.1% variation of the amplitude difference entails a $4\times10^{-13}$ frequency shift for a 20µT static magnetic field. With our working parameter values the magnetic shift is estimated to $1\times10^{-14}$ for a $2\times10^{-6}$ stability of the magnetic field. This sensitivity can be relaxed by working at lower magnetic field.

The temperature dependence of the buffer-gas pressure shift can be greatly reduced by a proper choice of the composition of the mixture of gases. In the absence of change of this composition or of the total pressure, the shift should be limited to the $10^{-14}$ level for a 1mK temperature variation. It can be reduced by decreasing the total pressure at the expense of the coherence lifetime, and consequently of the short-term stability, unless a wall coating is used.

A preliminary experimental study of the light power effects has shown that the shift rate depends on the total light power, on the power of each laser and on the cell temperature. This shift is not a pure AC Stark shift and requires a further detailed study. There are couples of values (laser intensity, temperature) for which the shift rate cancels. Nevertheless the shift is much more sensitive to a variation of the intensity of one laser ($4\times10^{-10}$ (mW/cm$^2$)$^{-1}$) than to a variation of the total intensity ($4\times10^{-11}$ (mW/cm$^2$)$^{-1}$). At constant intensity a variation of temperature can result in change of the shift rate and hence a frequency shift greater than the shift due to the pressure shift.

If the mean laser frequency is stabilized to within 10 kHz the shift is less than $10^{-14}$. The shift due to a variation of the optical detuning caused by a 1mK variation of the cell temperature (pressure shift of the optical transition) is negligible.

The main source of long term variations of the clock frequency appears then to be linked to the stability of laser intensities and polarizations. The polarization sensitivity is specific to the double Λ scheme and has no equivalent in a double-resonance clock. The laser intensity shift is increased because of the use of two independent lasers. It could maybe be reduced by using a single modulated laser as in push-pull scheme [42, 43]. But, either with two separate laser sources or with a single modulated laser source, it is preferable to stabilize power on each frequency than the total power. Nevertheless further studies are needed for a better understanding of the phenomena involved in the intensity shift in order to improve the long term stability of CPT clocks.

ACKNOWLEDGMENT

We are pleased to acknowledge Pierre Bonnay and Annie Gérard for manufacturing Cs cells with great skill. We are grateful to Michel Lours, José Pinto and Laurent Volodimer of the SYRTE electronic team for their helpful assistance. We are also thankful to Rodolphe Boudot for careful reading of the manuscript. We thank the anonymous reviewers for their comments and suggestions. This work was supported in part by DGA under Grant 2009.34.0052, and by LNE. O. K. acknowledges support from DGA.

REFERENCES

1. J. Camparo, "The rubidium atomic clock and basic research", *Physics Today*, pp. 33-39, nov. 2007.
2. J. Vanier and C. Audoin, *The quantum physics of atomic frequency standards*, IOP Publishing, Bristol and Philadelphia, 1989.
3. J. C. Camparo, "A partial analysis of drift in the rubidium gas cell atomic frequency standard", in *Proc. 18th Annual Precise Time and Time Interval (PTTI) Applications and Planning Meeting*, Washington D.C., USA, 1987, pp. 565-588.
4. J. Camparo, "Does the light shift drive frequency aging in the rubidium atomic clock?", *IEEE Trans. Ultrasonics Ferroelectr. Frequency Control*, vol.52, 1075, 2005.
5. J. C. Camparo, "Influence of the atmosphere on a rubidium clock's frequency aging", in *Proc. 39th Annual Precise Time and Time Interval (PTTI) Applications and Planning Meeting*, Long Beach, USA, 2007, pp. 317-322.

6. J. C. Camparo, C. M. Klimcak and S. J. Herbulok, "Frequency equilibration in the vapor-cell atomic clock", *IEEE Trans. Instrum. Meas.*, vol. 54, pp. 1873-1880, 2005.
7. G. Mileti, J. Q. Deng, F. L. Walls, D. A. Jennings, R. E. Drullinger, "Laser pumped rubidium frequency standards: new analysis and progress", *IEEE J. Quantum Electron.*, vol. **QE-34**, pp. 233-237, 1998.
8. C. Affolderbach, F. Droz, and G. Mileti, "Experimental demonstration of a compact and high-performance laser-pumped rubidium gas cell atomic frequency standard", *IEEE Trans. Instrum. Meas.*, vol. 55, pp. 429-435, 2006.
9. J. Vanier, C. Mandache, "The passive optically pumped Rb frequency standard: the laser approach", *Appl. Phys. B*, vol. 87, pp. 565-593, 2007.
10. T. Bandi et al., "Laser pumped high performance compact gas-cell Rb standard with $<3 \times 10^{-13} \tau^{-1/2}$ stability", in *Proc. 2012 European Frequency and Time Forum(EFTF)*, Gothenburg, 2012, pp. 494-496..
11. A. Godone, S. Micalizio, F. Levi, and C. Calosso, "Physics characterization and frequency stability of the pulsed rubidium maser", *Phys. Rev. A*, vol. 74, 043401, 2006.
12. S. Micalizio, C. E. Calosso, A. Godone and F. Levi, "Metrological characterization of the pulsed Rb clock with optical detection", *Metrologia*, vol. 49, 425, 2012.
13. E. Arimondo, "Coherent Population Trapping in Laser Spectroscopy", *Progress in Optics*, vol. XXXV, pp. 257-353, 1996.
14. N. Cyr, M. Têtu and M. Breton, All optical microwave frequency standard: a proposal, *IEEE Trans. Instrum. Meas.* Vol. 42, pp. 640, 1993.
15. J. Vanier, "Atomic clocks based on coherent population trapping: a review", *Applied Physics B*, vol. 81, pp. 421-442, 2005.
16. J. Kitching, S. Knappe, and L. Hollberg, "Miniature vapour-cell atomic-frequency references", *Appl. Phys. Lett.*, vol. 81, pp. 553-555, 2002.
17. S. Knappe, P. D. D. Schwindt, V. Shah, L. Hollberg, J. Kitching, L. Liew, and J Moreland, "A chip-scale atomic clock based on $^{87}$Rb with improved frequency stability", *Opt. Expr.*, vol. 13, pp. 1249-1253, 2005.
18. M. Zhu, "High contrast signal in coherent population trapping based atomic frequency standard application", in *Proc.2013 IEEE International Frequency Control Symposium (FCS) jointly with 17th European Frequency and Time Forum (EFTF)*, 2003, pp. 16-21.
19. A. Godone, F. Levi, S. Micalizio, C. Calosso, "Coherent population trapping maser: Noise spectrum and frequency stability", *Phys. Rev. A*, vol. 70, 012508, 2004.
20. T. Zanon., S. Guérandel, E. de Clercq, D. Holleville, N. Dimarcq, and A. Clairon, "High Contrast Ramsey Fringes with Coherent Population Trapping Pulses in a Double Lambda Atomic System", *Phys. Rev. Lett.*, vol. 94, 193002, 2005.
21. R. Boudot, S. Guérandel, E. de Clercq, N. Dimarcq, A. Clairon, "Current Status of a pulsed CPT Cs Cell Clock,", *IEEE Trans. Instr. Meas.*, vol. 58, pp.1217-1222, 2009.
22. S. Micalizio, A. Godone, F. Levi, and C. Calosso, "Medium-long term frequency stability of pulsed vapour cell clocks", *IEEE Trans. Ultrasonics Ferroelectr. Frequency Control*, vol. 57, pp. 1524-1534, 2010.
23. C. E. Calosso, A. Godone, F. Levi, and S. Micalizio, "Enhanced temperature sensitivity in vapour-cell frequency standards", *IEEE Trans. Ultrasonics Ferroelectr. Frequency Control*, vol. 59, pp. 2646-2654, 2012.
24. R. Boudot, S. Guérandel, E. de Clercq, " Simple-design low-noise NLTL-based frequency synthesizers for a CPT Cs Clock", *IEEE Trans. Instr. Meas.*, vol. 58, pp. 3659-3665, 2009.
25. S. Guérandel, T. Zanon, N. Castagna, F. Dahes, E. de Clercq, N. Dimarcq, and A. Clairon, "Raman-Ramsey interaction for coherent population trapping Cs clock", *IEEE Trans. Instr. Meas.*, vol. 56, pp.383-387, 2007.
26. N. Ramsey, *Molecular Beams*, Oxford University Press, New York, 1956.
27. J. Vanier, R. Kunski, N. Cyr, J. Y. Savard, and M. Têtu, "On hyperfine frequency shifts caused by buffer gases: Application to the optically pumped passive rubidium frequency standard", *J. Appl. Phys.*, vol. 53, 5387, 1982.
28. O. Kozlova, S. Guérandel, E. de Clercq, "Temperature and pressure shift of the Cs clock transition in presence of buffer gases Ne, N$_2$ Ar", *Phys. Rev. A*, vol. 83, 062714, 2011.
29. S. Knappe, J. Kitching, L. Hollberg, R. Wynands "Temperature dependence of coherent population trapping resonances", *Appl. Phys. B*, vol. 74, pp. 217-222, 2002.
30. M. Huang, J. G. Coffer, and J. C. Camparo, "$^{87}$Rb hyperfine-transition dephasing in mixed buffer-gas systems", *Phys. Rev. A,* vol. 75, pp. 052717-1-8, 2007.
31. S. Knappe, R. Wynands, J. Kitching, H. G. Robinson, and L. Hollberg, "Characterization of coherent population-trapping resonances as atomic frequency references", *J. Opt. Soc. Am. B*, vol. 18, pp. 1545-1553, 2001.
32. D. Miletic, C. Affolderbach, M. Hasegawa, R. Boudot, C. Gorecki, G. Mileti, "AC Stark-shift in CPT based Cs miniature atomic clocks", *Appl. Phys. B*, vol. 109, pp. 89-97, 2012.
33. J. P. Barrat and C. Cohen-Tannoudji, "Elargissement et déplacement des raies de resonance magnétique causes par une excitation optique", *J. Phys. Rad.*, vol. 22, pp. 443-450, 1961.
34. E. de Clercq and P. Cerez, "Evaluation of the light shift in a frequency standard based on Raman induced Ramsey resonance", *Opt. Comm.*, vol. 45, pp. 91-94, 1983.
35. J. Vanier, A. Godone, and F. Levi, "Coherent population trapping in cesium: dark lines and coherent microwave emission", *Phys. Rev. A*, vol. 58, pp. 2345-2358, 1998.
36. F. Levi, A. Godone, and J. Vanier, "The light-shift effect in the coherent population trapping cesium maser", *IEEE Trans. Ultrasonics Ferroelectr. Frequency Control*, vol. 47, pp. 466-470, 2000.
37. M Zhu, and L. S. Cutler, in *Proc. 32th Annual Precise Time and Time Interval (PTTI) Applications and Planning Meeting*, Reston, USA, 2007, pp. 311-324.
38. Zanon T., de Clercq E., Arimondo E., "Ultra-high resolution spectroscopy with atomic or molecular dark resonances: exact steady-state line shapes and asymptotic profiles in the pulsed adiabatic regime", *Phys. Rev. A*, vol. 84, 062502, 2011.
39. O. Kozlova, PhD Thesis, Université Pierre et Marie Curie, Paris, France, 2012.
40. J. C. Camparo, R. P. Frueholz, and C. H. Volk, "Inhomogeneous light shift in alkali-metal atoms", *Phys. Rev. A*, vol. 27, pp. 1914-1924, 1983.
41. X. Liu and R. Boudot, "A distributed-feedback diode laser frequency stabilized on Doppler-free Cs $D_1$ line", *IEEE Trans. Instr. Meas.*, vol. 61, pp. 2852-2855, 2012.
42. Y. Y. Jau, E. Miron, A. B. Post, N. N. Kuzma, and W. Happer, " Push-Pull optical pumping of pure superposition state", *Phys. Rev. Lett.*, vol. 93, 160802, 2004.
43. X. Liu, J-M. Mérolla, S. Guérandel, E. de Clercq, and R. Boudot, "Ramsey spectroscopy of high contrast Ramsey resonances with push-pull optical pumping in Cs vapor", *Opt. Expr.*, vol. 21, pp. 12451-12459, 2013.